\documentclass[10pt,twocolumn,letterpaper]{article}

\usepackage{cvpr} % uncomment this for the final submission
\usepackage{times}
\usepackage{epsfig}
\usepackage{graphicx}
\usepackage{amsmath}
\usepackage{amssymb}

\usepackage{tabularx}
\usepackage{array}
\usepackage{booktabs}
\usepackage{multirow}

\usepackage{tikz}
\usetikzlibrary{positioning,calc,fit,arrows.meta,shapes.geometric}
\usepackage{xcolor}

\begin{document}

%%%%%%%%% TITLE
\title{Shared-Private Cross-Modal Discrepancy Learning for Imbalanced Multi-Source Audio-Visual Deepfake Detection
}

\author{ Qiwu Wen, Jahangir Alam\\ Luqia Technologies\\ 405 Ogilvy Ave., Suite 101\\ Montréal, Québec H3N 1M3, Canada\\ {\tt\small \{qiwu.wen,jahangir.alam\}@luqia.ca} }

\maketitle
\thispagestyle{empty}

%%%%%%%%% ABSTRACT

\begin{abstract}
Audio-visual deepfake detectors are often trained under a single dataset or
manipulation regime, but multi-source training collections may contain
heterogeneous attacks and uneven source distributions. Training separate
detectors for each source is inefficient, while a single shared detector may
be biased toward dominant source-specific patterns. We study known-source
imbalanced multi-source audio-visual deepfake detection using FakeAVCeleb and
LAV-DF, where source identity is supplied during training and inference. To
address this setting, we build on the Multi-Scale Cross-Modal Transformer
Encoder (MSCT) and propose Pre-MSCT Shared-Private Feature Fusion. Our method routes
\texttt{[CLS]}-augmented audio and visual tokens into a shared stream and
source-specific private streams before the first MSCT block, while retaining
its internal multi-scale self-attention (MSSA), differential cross-modal
attention (DCA), and feed-forward operations. The shared stream models
cross-source reusable evidence,
while the private streams retain source-dependent residual variations. A
domain-aware gated residual fusion module injects selected private evidence
into the shared representation for the main fused prediction. Experiments
under two controlled source-composition protocols show improved macro and
worst-source AUC compared with the MRDF and MSCT baselines. These results support
shared-private cross-modal discrepancy learning in the evaluated two-source
setting with known source identity.
\end{abstract}

\section{Introduction}
\label{sec:introduction}

The rapid progress of generative models has made facial and speech
manipulation increasingly realistic, raising serious concerns for biometric
security, online identity verification, and media forensics. Early deepfake
detection methods mainly focused on either visual or audio artifacts in
manipulated faces~\cite{Dolhansky2020DFDC,Rossler2019FaceForensics} and
manipulated speech~\cite{add_alam_specom2021,add_chuchra_ijcb2025,add_Kang2022MixupRS,add_liu_ijcb2025,add_mengyu_ijcb2025},
but realistic attacks may involve both facial appearance and speech
signals~\cite{Astrid2024FineGrained,pavel_ijcb2023,ling_ijcb2025}. In
particular, audio-visual deepfakes can manipulate not only the visual stream
but also the spoken content, voice identity, or cross-modal synchronization.
This motivates audio-visual deepfake detection, where complementary cues from
facial dynamics, speech signals, and their temporal correspondence are jointly
exploited~\cite{Cai2023Glitch,Cai2022LAVDF,Khalid2021FakeAVCeleb}. In our
binary setting, a video is classified as fake if either its audio stream,
visual stream, or both are manipulated.

Recent audio-visual deepfake detectors commonly exploit cross-modal
inconsistency as an important cue. Contextual cross-modal attention has been
explored for audio-visual deepfake detection and localization, while neural
architecture search has been used to identify effective bimodal fusion
designs~\cite{katamneni2024contextual,pn2024straight}. For example,
Cross-Modality and Within-Modality Regularization for Audio-Visual DeepFake
Detection (MRDF) regularizes modality-specific and cross-modality features to
preserve useful audio-visual discrepancies~\cite{Zou2024MRDF}, while
feature-fusion and cross-modal learning methods aim to capture correspondence
between audio and visual streams~\cite{oorloff2024AVFF}. More recently, MSCT
strengthens cross-modal discrepancy modeling by combining MSSA with
DCA~\cite{Wei2026MSCT}. We use
MSCT as the backbone and leave its internal MSSA, DCA, and feed-forward
operations unchanged.

However, many existing audio-visual deepfake detectors are developed and
evaluated under relatively homogeneous or balanced benchmark settings, such
as training on a single source dataset or assuming balanced source
distributions. In multi-source training, datasets may vary in acquisition
protocol, compression level, face preprocessing, audio quality, manipulation
method, and synchronization characteristics. This heterogeneity is evident in
commonly used audio-visual deepfake benchmarks. FakeAVCeleb~\cite{Khalid2021FakeAVCeleb}
is organized around modality-defined audio/video manipulation categories,
including fake audio, fake video, and jointly manipulated audio-video samples.
In contrast, LAV-DF~\cite{Cai2022LAVDF,Cai2023Glitch} focuses on localized
temporal content-driven forgeries, where short audio, visual, or audio-visual
segments may be manipulated to alter the meaning of the video. Therefore, the
two datasets may exhibit different forgery cues: FakeAVCeleb emphasizes
modality-level manipulation patterns, while LAV-DF emphasizes temporally
localized semantic and synchronization inconsistencies.

This difference becomes more challenging under source imbalance. When one
source contributes substantially more training samples than another, a
detector may be biased toward the dominant source and may under-represent
forgery cues from the minority source. In such a setting, the challenge is not
only to detect audio-visual inconsistency, but also to determine which
inconsistency patterns should be shared across sources and which should be
treated as source-dependent residual variations.

This issue is particularly relevant for attention-based cross-modal
discrepancy models. While a single shared cross-modal pathway can be effective
in a homogeneous training setting, it may be suboptimal under imbalanced
multi-source training. For instance, in MSCT, the same DCA pathway is shared
by all samples~\cite{Wei2026MSCT}. When the source distribution is
heterogeneous and imbalanced, the shared attention-difference operator must
simultaneously capture transferable audio-visual forgery evidence and
source-dependent residual patterns. Reusable cross-source cues may therefore
become entangled with dataset-specific synchronization, acquisition, or
manipulation artifacts, especially when one source dominates training.

To address this problem, we propose a Pre-MSCT Shared-Private Feature Fusion
framework for known-source imbalanced multi-source audio-visual deepfake
detection. Instead of sending all samples through a single shared MSCT
pathway, our method routes \texttt{[CLS]}-augmented audio and visual tokens
into a shared stream and source-specific private streams before the first MSCT
block, i.e., before both MSSA and DCA. The shared stream is designed to model
cross-source reusable evidence, while the private streams retain
source-dependent residual variations. Under the known-source setting, the
source identity selects the private adapter, private MSCT expert, private
classifier, and a bounded source-specific modulation coefficient. A shared
feature-wise gate then injects the selected private representation into the
shared \texttt{[CLS]} representation through gated residual fusion. The fused
branch is used as the main prediction, while shared-only and private-only
branches are reported as diagnostic outputs.

We evaluate this setting on FakeAVCeleb and LAV-DF under controlled
source-sampling ratios. We compare our method with MRDF and MSCT
baselines, and analyze the shared-only, private-only, and fused branches to
understand the role of each component. Our study focuses on known-source
multi-source detection.

Our contributions are summarized as follows:
\begin{itemize}
    \item We formulate and study known-source imbalanced multi-source
    audio-visual deepfake detection, where heterogeneous source datasets
    contribute unevenly to training.
    \item We propose a Pre-MSCT shared-private framework that routes
    audio-visual tokens into shared and source-specific private streams before
    the first MSCT block.
    \item We introduce a domain-aware gated residual fusion module that
    injects selected private evidence into the shared representation while
    keeping the fused prediction as the main detection branch.
    \item We evaluate the proposed framework under controlled source
    compositions and compare it against MRDF and MSCT baselines, with
    diagnostic branch analysis and component ablations.
\end{itemize}

\section{Related work}
\label{sec:related_work}

\paragraph{Audio-visual deepfake detection.}
Visual deepfake detection has been extensively studied for identifying
artifacts in manipulated facial images and
videos~\cite{Rossler2019FaceForensics,Dolhansky2020DFDC}, while audio-visual
deepfake detection further exploits inconsistencies between facial dynamics,
speech content, voice characteristics, and cross-modal synchronization. Early
audio-visual methods jointly model audio and visual streams to capture
complementary manipulation cues~\cite{Zhou2021JointAV}, and audiovisual
representation learning has been used to improve multimodal trace modeling
for deepfake detection~\cite{Raza2023Multimodaltrace}. Recent methods further
study cross-modal discrepancy, feature fusion, and fine-grained inconsistency
modeling. MRDF regularizes cross-modality and within-modality representations
to preserve discrepancy cues~\cite{Zou2024MRDF}, AVFF studies audio-visual
feature fusion~\cite{oorloff2024AVFF}, and contextual cross-modal attention
captures localized inconsistency for detection and
localization~\cite{katamneni2024contextual}. Fine-grained spatial and temporal
inconsistency modeling has also been explored~\cite{Astrid2024FineGrained}.
In addition, bimodal neural architecture search has been used to identify
effective audio-visual fusion designs~\cite{pn2024straight}. MSCT further
strengthens discrepancy modeling through MSSA and DCA~\cite{Wei2026MSCT}.
These methods demonstrate the importance of audio-visual inconsistency, but
generally process samples through a unified pathway. In contrast, our work
focuses on imbalanced multi-source detection, where transferable evidence and
source-dependent residual cues may need to be modeled separately.

\paragraph{Audio-visual deepfake datasets and manipulation regimes.}
The development of audio-visual deepfake detection has been facilitated by
benchmarks incorporating diverse generation techniques. Notable examples
include FakeAVCeleb~\cite{Khalid2021FakeAVCeleb},
LAV-DF~\cite{Cai2022LAVDF,Cai2023Glitch}, TVIL~\cite{tvil2023},
AV-Deepfake1M and AV-Deepfake1M++~\cite{Cai2024AVDeepfake1M,avdeepfake1m++},
and SWAN-DF~\cite{pavel_ijcb2023}.

These datasets differ not only in scale and source content, but also in
manipulation regimes and resulting forgery cues. FakeAVCeleb is organized
around modality-defined categories, including fake visual streams, fake audio
streams, and jointly manipulated samples~\cite{Khalid2021FakeAVCeleb}. Such
manipulations can introduce cues related to facial identity replacement,
synthesized speech, voice conversion, or audio-video mismatch. LAV-DF focuses
on localized temporal content-driven forgeries, where short audio, visual, or
joint segments are manipulated to alter meaning~\cite{Cai2022LAVDF,Cai2023Glitch}.
Larger benchmarks further emphasize content-driven audio, visual, and joint
manipulations at scale~\cite{Cai2024AVDeepfake1M}. Thus, combining FakeAVCeleb
and LAV-DF forms a heterogeneous multi-source setting with different cue
distributions. More broadly, deepfake phylogeny and attribution studies
suggest that generation processes and manipulation histories may leave
source-dependent traces~\cite{narayan2022deephy}. Cross-dataset studies also
show degradation under distribution shift~\cite{Nadimpalli2022CrossDataset}.
These observations motivate separating reusable evidence from
source-dependent residual patterns in heterogeneous multi-source training.

\paragraph{Shared-private learning for heterogeneous sources.}
Shared-private representation learning is a common strategy for heterogeneous
data. Domain separation networks decompose representations into shared and
private components to separate domain-invariant information from
domain-specific factors~\cite{Bousmalis2016DomainSeparation}.
Domain-adversarial and domain-generalization methods also aim to improve
robustness under domain shifts~\cite{ganin2016domain,li2018deep}. In
audio-visual deepfake detection, modality-invariant and modality-specific
learning has been explored to preserve common and modality-dependent
cues~\cite{Katamneni2023MISAVoiDD}. Building on this established factorization,
our contribution is its source-specific pre-MSCT realization and gated
residual integration for known-source imbalanced audio-visual detection.
Instead of suppressing source-specific information, our method preserves
source-dependent residual evidence and injects it into a shared audio-visual
representation. Applying this decomposition before the first MSCT block lets
the two streams perform discrepancy modeling in distinct pathways.

\section{Method}
\label{sec:method}

\subsection{Problem formulation}
\label{sec:problem_formulation}
We study known-source imbalanced multi-source audio-visual deepfake detection.
Given a set of source datasets
\(\mathcal{D}=\{\mathcal{D}_d\}_{d=1}^{K}\), each sample is denoted as
\((x_a,x_v,y,d)\), where \(x_a\) and \(x_v\) are the audio and visual inputs,
\(y\in\{0,1\}\) is the binary label with \(y=1\) denoting a fake sample, and
\(d\in\{1,\dots,K\}\) is the source identity. In this work, \(d\) is
externally supplied during both training and known-source inference; the model
does not infer it. Our goal is to learn a detector that remains robust under
heterogeneous and imbalanced source distributions. Unknown-source or
unseen-domain inference is outside the scope of this work.

\subsection{MSCT backbone and motivation}
\label{sec:msct_backbone}

We build our method upon the MSCT backbone~\cite{Wei2026MSCT}. Audio is
represented by 26-dimensional log-filterbank features stacked in groups of
four, yielding 104-dimensional inputs that a linear layer projects to
\(D=256\). Video frames are encoded into 256-dimensional tokens by the
WTConv/Res2Net-CBAM visual front-end. The audio and visual tokens are then
aligned to a common sequence length. Additional preprocessing and front-end
details are provided in the supplementary material. Learnable classification
tokens are prepended to both modality sequences. Following common Transformer
terminology, we refer to these as \texttt{[CLS]} tokens and denote the
resulting sequences as
\begin{equation}
\begin{aligned}
X_a &\in \mathbb{R}^{(T+1)\times D}, \\
X_v &\in \mathbb{R}^{(T+1)\times D},
\end{aligned}
\label{eq:input_tokens}
\end{equation}
where \(T\) is the common token length and \(D=256\). Alignment uses
index-level padding and truncation rather than explicit timestamp matching.

MSCT processes the audio and visual sequences in Eq.~\eqref{eq:input_tokens}
using stacked MSCT blocks. Each block contains MSSA, DCA, and an FFN. DCA
computes differential attention by subtracting a cross-modal-query-conditioned
attention map from the intra-modal attention map for each modality. The final
audio and visual \texttt{[CLS]} tokens are concatenated for real/fake
classification.

Although MSCT provides an effective cross-modal discrepancy mechanism, it
uses a single shared MSCT pathway for all samples. Under imbalanced
multi-source training, the same DCA parameters must model both cross-source
reusable forgery evidence and source-dependent residual variations. This may
entangle transferable inconsistency cues with dataset-specific synchronization
patterns, acquisition characteristics, or artifacts. We therefore replace the
single pathway with a Pre-MSCT shared-private decomposition.

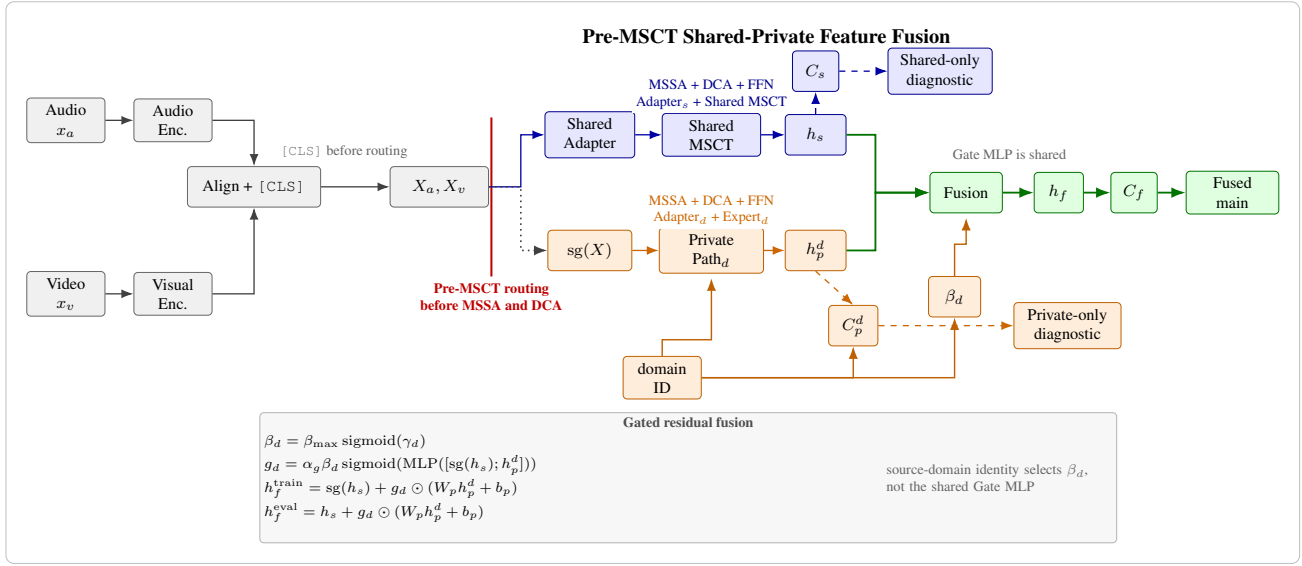
\begin{figure*}[t]
\centering
\resizebox{0.98\textwidth}{!}{%
\begin{tikzpicture}[
    x=1cm,y=1cm,
    font=\footnotesize,
    >=Latex,
    box/.style={rectangle,rounded corners=2pt,draw=black!60,
        line width=0.45pt,align=center,inner sep=3pt,
        minimum height=0.70cm,minimum width=1.35cm},
    graybox/.style={box,fill=black!6},
    sharedbox/.style={box,fill=blue!10,draw=blue!55!black},
    privatebox/.style={box,fill=orange!14,draw=orange!75!black},
    fusionbox/.style={box,fill=green!12,draw=green!45!black},
    redline/.style={draw=red!75!black,line width=1.0pt},
    normarrow/.style={->,line width=0.65pt,draw=black!75},
    mainarrow/.style={->,line width=0.90pt,draw=green!45!black},
    diagarrow/.style={->,dashed,line width=0.65pt,draw=black!65},
    stoparrow/.style={->,dotted,line width=0.80pt,draw=black!75},
    selectarrow/.style={->,line width=0.70pt,draw=orange!75!black}
]

\node[font=\bfseries] (figtitle) at (12.9,7.45)
{Pre-MSCT Shared-Private Feature Fusion};

\node[graybox] (audio) at (0.8,6.0) {Audio\\$x_a$};
\node[graybox] (video) at (0.8,3.0) {Video\\$x_v$};
\node[graybox] (aenc) at (2.65,6.0) {Audio\\Enc.};
\node[graybox] (venc) at (2.65,3.0) {Visual\\Enc.};
\node[graybox,minimum width=2.30cm] (align) at (4.05,4.85)
{Align + \texttt{[CLS]}};
\node[graybox,minimum width=1.70cm] (tokens) at (7.25,4.85) {$X_a,X_v$};

\draw[normarrow] (audio.east) -- (aenc.west);
\draw[normarrow] (video.east) -- (venc.west);
\coordinate (aelbow) at ($(aenc.east)+(0.70,0)$);
\coordinate (velbow) at ($(venc.east)+(0.70,0)$);
\draw[normarrow] (aenc.east) -- (aelbow) -| (align.north);
\draw[normarrow] (venc.east) -- (velbow) -| (align.south);
\draw[normarrow] (align.east) -- (tokens.west);
\node[font=\scriptsize,text=black!60,fill=white,inner sep=1pt]
at (5.60,5.45) {\texttt{[CLS]} before routing};

\draw[redline] (8.15,6.05) -- (8.15,3.30);
\node[font=\scriptsize\bfseries,text=red!75!black,align=center,
fill=white,inner sep=1pt] at (8.15,2.95)
{Pre-MSCT routing\\before MSSA and DCA};

\node[sharedbox,minimum width=1.55cm] (sad) at (9.85,5.75)
{Shared\\Adapter};
\node[sharedbox,minimum width=1.70cm] (smsct) at (11.95,5.75)
{Shared\\MSCT};
\node[sharedbox,minimum width=1.05cm] (hs) at (13.75,5.75) {$h_s$};
\draw[normarrow,draw=blue!65!black] (tokens.east) -- ++(0.55,0) |- (sad.west);
\draw[normarrow,draw=blue!65!black] (sad) -- (smsct);
\draw[normarrow,draw=blue!65!black] (smsct) -- (hs);
\node[font=\scriptsize,text=blue!55!black,align=center,
fill=white,inner sep=2pt] at (11.95,6.45)
{MSSA + DCA + FFN\\Adapter$_s$ + Shared MSCT};

\node[sharedbox,minimum width=0.80cm] (cs) at (13.75,6.85) {$C_s$};
\node[sharedbox,minimum width=1.80cm] (sout) at (15.90,6.85)
{Shared-only\\diagnostic};
\draw[diagarrow,draw=blue!65!black] (hs.north) -- (cs.south);
\draw[diagarrow,draw=blue!65!black] (cs) -- (sout);

\node[privatebox,minimum width=1.45cm] (sgx) at (9.85,3.75)
{$\mathrm{sg}(X)$};
\node[privatebox,minimum width=1.80cm] (ppath) at (11.95,3.75)
{Private\\Path$_d$};
\node[privatebox,minimum width=1.05cm] (hp) at (13.75,3.75) {$h_p^d$};
\draw[stoparrow] (tokens.east) -- ++(0.55,0) |- (sgx.west);
\draw[normarrow,draw=orange!80!black] (sgx) -- (ppath);
\draw[normarrow,draw=orange!80!black] (ppath) -- (hp);
\node[font=\scriptsize,text=orange!80!black,align=center,
fill=white,inner sep=2pt] at (11.95,4.45)
{MSSA + DCA + FFN\\Adapter$_d$ + Expert$_d$};

\node[privatebox,minimum width=1.35cm] (dom) at (11.10,1.55)
{domain\\ID};
\node[privatebox,minimum width=0.85cm] (cp) at (14.40,2.45) {$C_p^d$};
\node[privatebox,minimum width=0.90cm] (beta) at (16.15,2.95) {$\beta_d$};
\node[privatebox,minimum width=1.85cm] (pout) at (18.10,2.45)
{Private-only\\diagnostic};
\draw[diagarrow,draw=orange!80!black] (hp.south) -- (cp.north);
\draw[diagarrow,draw=orange!80!black] (cp) -- (pout);
\draw[selectarrow] (dom.north) -- ++(0,0.28) -| ([yshift=-0.08cm]ppath.south);
\coordinate (domsplit) at ($(dom.east)+(0.55,0)$);
\coordinate (cpup) at (cp.south |- domsplit);
\coordinate (betaturn) at (beta.south |- domsplit);
\draw[selectarrow] (dom.east) -- (domsplit) -- (cpup) -- (cp.south);
\draw[selectarrow] (dom.east) -- (domsplit) -- (betaturn) -- (beta.south);

\node[fusionbox,minimum width=1.25cm] (fuse) at (16.35,4.75) {Fusion};
\node[fusionbox,minimum width=0.85cm] (hf) at (17.95,4.75) {$h_f$};
\node[fusionbox,minimum width=0.80cm] (cf) at (19.25,4.75) {$C_f$};
\node[fusionbox,minimum width=1.60cm] (fout) at (20.95,4.75)
{Fused\\main};
\draw[mainarrow] (hs.east) -- ++(0.50,0) |- (fuse.west);
\draw[mainarrow] (hp.east) -- ++(0.50,0) |- (fuse.west);
\draw[mainarrow] (fuse) -- (hf);
\draw[mainarrow] (hf) -- (cf);
\draw[mainarrow] (cf) -- (fout);
\draw[selectarrow] (beta.north) -- ++(0,0.55) -| ([yshift=-0.08cm]fuse.south);
\node[font=\scriptsize,text=black!60,fill=white,inner sep=1pt]
at (17.10,5.42) {Gate MLP is shared};

\node[draw=black!30,rounded corners=2pt,fill=black!3,
minimum width=14.8cm,minimum height=2.25cm] (formula) at (11.55,-0.18) {};
\node[font=\scriptsize\bfseries,text=black!70,fill=black!3,
inner sep=1pt] at (11.55,0.78) {Gated residual fusion};
\node[anchor=west,align=left,font=\scriptsize] at (4.10,-0.18) {%
\begin{tabular}{@{}l@{}}
\(\beta_d=\beta_{\max}\,\mathrm{sigmoid}(\gamma_d)\)\\[1mm]
\(g_d=\alpha_g\beta_d\,\mathrm{sigmoid}(\mathrm{MLP}([\mathrm{sg}(h_s);h_p^d]))\)\\[1mm]
\(h_f^{\mathrm{train}}=\mathrm{sg}(h_s)+g_d\odot(W_p h_p^d+b_p)\)\\[1mm]
\(h_f^{\mathrm{eval}}=h_s+g_d\odot(W_p h_p^d+b_p)\)
\end{tabular}};
\node[anchor=west,align=left,font=\scriptsize,text=black!60,
text width=4.7cm] at (14.85,-0.18)
{source-domain identity selects \(\beta_d\),\\not the shared Gate MLP};

\node[draw=black!25,rounded corners=4pt,
fit=(audio)(video)(fout)(sout)(pout)(formula)(figtitle),
inner xsep=10pt,inner ysep=10pt] (panel) {};
\end{tikzpicture}%
}
\vspace{-1mm}
\caption{Pre-MSCT Shared-Private Feature Fusion. Tokens are routed before MSSA
and DCA. Source identity selects the private path and \(\beta_d\), while the
gate MLP is shared. Dashed and dotted arrows denote diagnostic and
stop-gradient paths, respectively.}
\label{fig:method}
\end{figure*}

Figure~\ref{fig:method} illustrates the proposed pipeline, and
Table~\ref{tab:method_comparison} summarizes its architectural differences
from MSCT. Our framework introduces shared/private token routing before the
first MSCT block and feature-level fusion of the selected private evidence,
while retaining the original MSSA and DCA operations.
Since DCA explicitly models audio-visual discrepancy, pre-MSCT routing lets
shared and private streams extract discrepancy cues in distinct pathways.

\begin{table*}[t]
\centering
\caption{Architectural comparison between MSCT and our Pre-MSCT
shared-private framework.}
\label{tab:method_comparison}
\small
\setlength{\tabcolsep}{5pt}
\renewcommand{\arraystretch}{1.12}
\begin{tabularx}{\textwidth}{|p{0.20\textwidth}|X|X|}
\hline
\textbf{Component} & \textbf{MSCT} & \textbf{Ours} \\
\hline
Token routing
& Unified pathway
& Shared/private routing before the first MSCT block \\
\hline
Cross-modal modeling
& Single shared MSCT stack
& Shared stack + source-specific private stacks with the same MSCT blocks \\
\hline
Private branch
& Not used
& Source-specific adapter, expert, and classifier \\
\hline
Domain identity
& Not used
& Selects private adapter, expert, \(C_p^d\), and \(\beta_d\) \\
\hline
Final representation
& Concatenated audio/video \texttt{[CLS]}
& Gated private residual on shared \texttt{[CLS]} representation \\
\hline
Gate design
& Not used
& Shared gate with bounded source modulation \\
\hline
Main output
& Unified classifier
& Fused classifier \\
\hline
Diagnostic outputs
& No shared/private diagnostics
& Shared-only and private-only \\
\hline
\end{tabularx}
\end{table*}

\subsection{Pre-MSCT shared-private token routing}
\label{sec:token_routing}

The residual adapters operate on complete audio and visual token sequences
after the learnable \texttt{[CLS]} tokens are prepended. Because routing
precedes the first shared or private MSCT block, both MSSA and DCA operate on
the routed representations and model discrepancy in distinct feature spaces.

\paragraph{Shared stream.}
The shared stream is designed to model cross-source reusable evidence. Given
\(X_a\) and \(X_v\), we apply residual adapters to obtain
\begin{equation}
\begin{aligned}
X_a^s &= X_a + \alpha_a U_a^s \phi(D_a^s X_a), \\
X_v^s &= X_v + \alpha_a U_v^s \phi(D_v^s X_v),
\end{aligned}
\label{eq:shared_adapters}
\end{equation}
where \(D^s\) and \(U^s\) denote the down-projection and up-projection layers
of the shared adapters, \(\phi(\cdot)\) is a non-linear activation, and
\(\alpha_a\) is a fixed adapter scaling hyperparameter. The adapted shared
tokens are then processed by the shared MSCT stack:
\begin{equation}
(H_a^s,H_v^s)=\mathrm{MSCT}_s(X_a^s,X_v^s).
\label{eq:shared_msct}
\end{equation}
Each shared MSCT block still contains MSSA, DCA, and FFN layers. The shared
representation is obtained by concatenating the final audio and visual
\texttt{[CLS]} outputs:
\begin{equation}
h_s=[H_a^{s,[\mathrm{CLS}]};H_v^{s,[\mathrm{CLS}]}].
\label{eq:shared_repr}
\end{equation}

\paragraph{Private stream.}
For each source domain \(d\), we introduce a domain-specific private adapter
and a private MSCT expert. The private stream is encouraged to retain
source-dependent residual variations, which may include synchronization
patterns, acquisition characteristics, or dataset-specific artifacts. We do
not assume that the private representation corresponds to purely semantic
forgery factors.

To isolate the shared pathway from gradients introduced by private-specific
objectives, the private adapters receive detached token sequences:
\begin{equation}
\begin{aligned}
X_a^{p,d}
&=\mathrm{sg}(X_a)
+\alpha_a U_a^{p,d}\phi(D_a^{p,d}\mathrm{sg}(X_a)), \\
X_v^{p,d}
&=\mathrm{sg}(X_v)
+\alpha_a U_v^{p,d}\phi(D_v^{p,d}\mathrm{sg}(X_v)),
\end{aligned}
\label{eq:private_adapters}
\end{equation}
where \(\mathrm{sg}(\cdot)\) denotes stop-gradient. The selected private expert
is determined by source identity \(d\):
\begin{equation}
(H_a^{p,d},H_v^{p,d})
=\mathrm{MSCT}_p^d(X_a^{p,d},X_v^{p,d}).
\label{eq:private_msct}
\end{equation}
Each private expert is a complete MSCT stack with MSSA, DCA, and FFN layers,
but its parameters are not shared across source domains. The private
representation is defined as
\begin{equation}
h_p^d=[H_a^{p,d,[\mathrm{CLS}]};H_v^{p,d,[\mathrm{CLS}]}].
\label{eq:private_repr}
\end{equation}

\subsection{Shared gate with bounded domain-specific modulation}
\label{sec:gated_fusion}

We fuse the shared and private representations in
Eqs.~\eqref{eq:shared_repr} and~\eqref{eq:private_repr} at the
\texttt{[CLS]} feature level rather than at the logit level. The fusion module
uses a shared feature-wise gate modulated by a bounded source-specific
coefficient. Specifically, we define
\begin{equation}
\beta_d=\beta_{\max}\cdot\mathrm{sigmoid}(\gamma_d),
\label{eq:beta}
\end{equation}
where \(\gamma_d\) is a learnable source-specific scalar and
\(\beta_{\max}\) bounds the modulation coefficient. The feature-wise gate is
computed as
\begin{equation}
g_d
=\alpha_g\beta_d\cdot\mathrm{sigmoid}
\left(\mathrm{MLP}([\mathrm{sg}(h_s);h_p^d])\right),
\label{eq:gate}
\end{equation}
where \(\alpha_g\) is a fixed global gate scaling hyperparameter. As shown in
Eq.~\eqref{eq:gate}, the gate MLP is shared across all source domains; source
identity affects fusion through the selected private feature \(h_p^d\) and
coefficient \(\beta_d\).

During training, the fused feature is computed as
\begin{equation}
h_{\mathrm{fused}}^{\mathrm{train}}
=\mathrm{sg}(h_s)+g_d\odot(W_p h_p^d+b_p),
\label{eq:fused_train}
\end{equation}
where \(W_p\) and \(b_p\) are the private-to-fused projection parameters, and
\(\odot\) denotes element-wise multiplication. The stop-gradient operation in
Eq.~\eqref{eq:fused_train} isolates the shared residual path from gradients
introduced by the fused objective. During known-source inference, we use the
undetached shared representation:
\begin{equation}
h_{\mathrm{fused}}^{\mathrm{eval}}
=h_s+g_d\odot(W_p h_p^d+b_p).
\label{eq:fused_eval}
\end{equation}
The stop-gradient operator does not alter forward values.
Equations~\eqref{eq:fused_train} and~\eqref{eq:fused_eval} therefore define
the same numerical feature; the separate notation only exposes the
training-time gradient routing.

\subsection{Prediction heads}
\label{sec:prediction_heads}

The model produces three prediction branches. We denote the shared and private
classifier logits as
\begin{equation}
\begin{aligned}
\ell_s &= C_s(h_s), \\
\ell_p^d &= C_p^d(h_p^d),
\end{aligned}
\label{eq:shared_private_logits}
\end{equation}
where \(C_s\) is the shared classifier and \(C_p^d\) is the selected
domain-specific private classifier. The fused logits are defined as
\begin{equation}
\begin{aligned}
\ell_f^{\mathrm{train}}
&=C_f(h_{\mathrm{fused}}^{\mathrm{train}}), \\
\ell_f^{\mathrm{eval}}
&=C_f(h_{\mathrm{fused}}^{\mathrm{eval}}),
\end{aligned}
\label{eq:fused_logits}
\end{equation}
where \(C_f\) is the fused classifier. The corresponding prediction
probabilities are obtained by applying the sigmoid function. At inference,
the main fused prediction probability is
\begin{equation}
p_f=\mathrm{sigmoid}(\ell_f^{\mathrm{eval}}).
\label{eq:fused_probability}
\end{equation}
The fused prediction is the main branch; shared-only and private-only outputs
are diagnostic ablations.

\subsection{Training objective}
\label{sec:training_objective}

The overall training objective consists of prediction losses and
representation regularizers:
\begin{equation}
\mathcal{L}=\mathcal{L}_{\mathrm{pred}}+\mathcal{L}_{\mathrm{reg}},
\label{eq:overall_loss}
\end{equation}
where the prediction loss is
\begin{equation}
\mathcal{L}_{\mathrm{pred}}
=\mathcal{L}_{\mathrm{shared}}
+\lambda_f\mathcal{L}_{\mathrm{fused}}
+\lambda_p\mathcal{L}_{\mathrm{private}},
\label{eq:prediction_loss}
\end{equation}
and the regularization loss is
\begin{equation}
\begin{aligned}
\mathcal{L}_{\mathrm{reg}}
=&\lambda_o\mathcal{L}_{\mathrm{orth}}
+\lambda_v\mathcal{L}_{\mathrm{var}} \\
&+\lambda_\beta\mathcal{L}_{\beta}
+\lambda_d\mathcal{L}_{\mathrm{domain}}.
\end{aligned}
\label{eq:regularization_loss}
\end{equation}

The prediction losses supervise the three output branches:
\begin{equation}
\begin{aligned}
\mathcal{L}_{\mathrm{shared}}
&=\mathrm{BCEWithLogits}(\ell_s,y), \\
\mathcal{L}_{\mathrm{fused}}
&=\mathrm{BCEWithLogits}(\ell_f^{\mathrm{train}},y), \\
\mathcal{L}_{\mathrm{private}}
&=\mathrm{BCEWithLogits}(\ell_p^d,y).
\end{aligned}
\label{eq:branch_prediction_losses}
\end{equation}
Here, \(\mathcal{L}_{\mathrm{fused}}\) optimizes the main fused branch during
training, while \(\mathcal{L}_{\mathrm{shared}}\) and
\(\mathcal{L}_{\mathrm{private}}\) keep the diagnostic branches independently
discriminative.

The representation regularizers are applied to the final concatenated
audio-visual \texttt{[CLS]} representations rather than intermediate DCA
features. The orthogonality regularizer encourages non-redundant shared and
private features:
\begin{equation}
\mathcal{L}_{\mathrm{orth}}
=\mathrm{Orth}(\mathrm{sg}(h_s),h_p^d).
\label{eq:orth_loss}
\end{equation}
We include a variance-floor regularizer for the private representation:
\begin{equation}
\mathcal{L}_{\mathrm{var}}=\mathrm{VarFloor}(h_p^d).
\label{eq:var_loss}
\end{equation}
The beta regularizer discourages excessive reliance on source-specific gated
corrections:
\begin{equation}
\mathcal{L}_{\beta}=\sum_{d=1}^{K}\beta_d^2.
\label{eq:beta_loss}
\end{equation}
Finally, the domain loss encourages the private representation to retain
source-dependent information:
\begin{equation}
\mathcal{L}_{\mathrm{domain}}=\mathrm{CE}(C_d(h_p^d),d),
\label{eq:domain_loss}
\end{equation}
where \(C_d\) is a domain classifier. These regularizers encourage
complementary shared and private representations, but they do not guarantee
complete disentanglement or eliminate dataset-specific shortcuts.

\subsection{Optimization and known-source inference}
\label{sec:optimization_inference}

In our main experiments, we employ a one-epoch shared-stream warm-up. During
this stage, private computation is disabled and the fused/private losses and
private regularizers are set to zero; the shared branch is optimized alone.
After warm-up, all configured objective terms are enabled while stop-gradient
continues to isolate the shared pathway from private-specific and
fused-objective gradients.

For fused and private inference, source identity selects the private adapter,
private MSCT expert, private classifier, and \(\beta_d\). The fused branch
remains the main prediction branch.

\section{Experiments}
\label{sec:experiments}

\begin{table}[t]
\centering
\caption{Source-sampling protocols; ratios are enforced over samples per
epoch, not per mini-batch.}
\label{tab:sampling_protocol}
\small
\setlength{\tabcolsep}{2.8pt}
\renewcommand{\arraystretch}{1.10}
\begin{tabular}{lccc}
\toprule
\textbf{Setting} & \textbf{Pool (FAC/LAV-DF)} & \textbf{Samples/epoch} & \textbf{Ratio} 
\\
\midrule
R13 & 320 / 960 & 1,280 & 1:3 \\
R19 & 320 / 2,880 & 3,200 & 1:9 \\
\bottomrule
\end{tabular}
\end{table}

\begin{table*}[!t]
\centering
\caption{Source-domain evaluation under two sampling ratios. FakeAVCeleb uses
its evaluation pool and LAV-DF its official test set. Higher AUC/ACC and lower
\(\Delta\) are better; per-ratio best results are bold. Aggregate metrics are
computed before rounding.}
\label{tab:main_source_comparison}
\small
\setlength{\tabcolsep}{4.5pt}
\renewcommand{\arraystretch}{1.10}
\begin{tabular}{llcccccc}
\toprule
\textbf{Ratio} & \textbf{Method}
& \textbf{FAC AUC} & \textbf{LAV-DF AUC}
& \textbf{Macro AUC} & \textbf{Macro ACC}
& \textbf{Worst AUC} & \textbf{\(\Delta\)AUC} \\
\midrule
\multirow{4}{*}{1:3}
& MRDF-CE & 0.8101 & 0.5116 & 0.6608 & 0.5127 & 0.5116 & 0.2985 \\
& MRDF-Margin & 0.8263 & 0.5058 & 0.6661 & 0.5049 & 0.5058 & 0.3205 \\
& MSCT & 0.9028 & \textbf{0.9564} & 0.9296 & 0.8503 & 0.9028 & 0.0537 \\
& \textbf{Ours} & \textbf{0.9565} & 0.9557 & \textbf{0.9561} & \textbf{0.9047} & \textbf{0.9557} & \textbf{0.0008} \\
\midrule
\multirow{4}{*}{1:9}
& MRDF-CE & 0.6991 & 0.4958 & 0.5974 & 0.7329 & 0.4958 & 0.2034 \\
& MRDF-Margin & 0.7531 & 0.5044 & 0.6288 & 0.7427 & 0.5044 & 0.2488 \\
& MSCT & 0.8911 & 0.9773 & 0.9342 & 0.8586 & 0.8911 & 0.0863 \\
& \textbf{Ours} & \textbf{0.9486} & \textbf{0.9843} & \textbf{0.9664} & \textbf{0.9174} & \textbf{0.9486} & \textbf{0.0358} \\
\bottomrule
\end{tabular}
\end{table*}

\subsection{Experimental setup}
\label{sec:experimental_setup}

\paragraph{Datasets.}
We use FakeAVCeleb~\cite{Khalid2021FakeAVCeleb} and
LAV-DF~\cite{Cai2022LAVDF,Cai2023Glitch} as source datasets, whose manipulation
characteristics are discussed in Sec.~\ref{sec:introduction}. Validation data
from the source domains is used for checkpoint selection. Final evaluation is
conducted on the official LAV-DF test split and on a held-out category-balanced
FakeAVCeleb split with 100 samples per manipulation category.
\paragraph{Metrics.}
The area under the receiver operating characteristic curve (AUC) summarizes
the detector's ability to rank fake samples above real samples across decision
thresholds. We report source-domain AUC, macro AUC, macro accuracy, worst-source AUC, and
source-domain AUC difference. Macro AUC is the primary ranking metric because
it is threshold-independent and less sensitive to source imbalance. For a
metric \(m\), such as AUC or accuracy, let \(m_{\mathrm{FAC}}\) and
\(m_{\mathrm{LAVDF}}\) denote values on the FakeAVCeleb evaluation pool and
LAV-DF test set, respectively. We define
\begin{equation}
\begin{aligned}
\mathrm{Macro}(m)&=\frac{m_{\mathrm{FAC}}+m_{\mathrm{LAVDF}}}{2},\\
\mathrm{Worst}(m)&=\min(m_{\mathrm{FAC}},m_{\mathrm{LAVDF}}),\\
\Delta(m)&=|m_{\mathrm{FAC}}-m_{\mathrm{LAVDF}}|.
\end{aligned}
\label{eq:metrics}
\end{equation}
Higher macro and worst-source performance is better, while lower \(\Delta\)
indicates more balanced source-domain performance. Accuracy uses a fixed
probability threshold of 0.5.

\paragraph{Source-sampling protocols.}
To evaluate robustness under source imbalance, we construct controlled
source-sampling protocols with different FakeAVCeleb:LAV-DF ratios.
Table~\ref{tab:sampling_protocol} summarizes the training pools and sampling
strategy. Weighted sampling is performed with replacement. Source ratios are
enforced in expectation at the epoch level rather than guaranteed per
mini-batch. When class balancing is enabled, sampling weights also balance
real/fake samples in expectation within each source.

Because R13 and R19 differ in source ratio, LAV-DF pool size, and samples per
epoch, we treat them as separate source-composition protocols and compare
methods within each protocol.

\paragraph{Baselines and implementation details.}
We compare with MRDF using cross-entropy supervision
(MRDF-CE)~\cite{Zou2024MRDF}, MRDF using a margin-based objective
(MRDF-Margin)~\cite{Zou2024MRDF}, and MSCT~\cite{Wei2026MSCT}. All methods use
identical fixed data manifests. MSCT and Ours share the linear audio token
encoder and WTConv/Res2Net-CBAM visual front-end; MRDF uses the audio and
visual feature extractors from the reproduced \texttt{AVHubertModel}
implementation. The MSCT baseline follows its
multimodal, unimodal, and alignment objectives, whereas these auxiliary
unimodal and alignment losses are disabled for Ours. Checkpoints are selected
by validation macro AUC of the main output. All models use Adam, batch size one
with four-step gradient accumulation, 16-bit mixed precision, at most 30
epochs, early-stopping patience 10, and weight decay \(10^{-4}\). The learning
rate is \(2\times10^{-4}\) for MRDF and \(10^{-4}\) for MSCT and Ours. For
Ours, the fused branch is main, and diagnostic branches are evaluated from the
same fused-selected checkpoint. Additional preprocessing and architecture
hyperparameters are provided in the supplementary material.

\subsection{Main results under source imbalance}
\label{sec:main_results}

Table~\ref{tab:main_source_comparison} reports the main source-domain results
under the 1:3 and 1:9 protocols. We report the fused branch of Ours as the main
output. Compared with the MRDF and MSCT baselines, Ours improves macro AUC
under both protocols. In 1:3, it improves macro AUC from 0.9296 to 0.9561 over
MSCT and reduces the source difference from 0.0537 to 0.0008. In 1:9, it
improves macro AUC from 0.9342 to 0.9664 and reduces the difference from
0.0863 to 0.0358. These results support the effectiveness of Pre-MSCT
shared-private routing and gated residual fusion.

Under R13, Ours raises FakeAVCeleb AUC from 0.9028 to 0.9565 while maintaining
comparable LAV-DF AUC, yielding nearly equal source performance. Under R19, it
improves both source AUCs over MSCT, so the macro gain does not come at the
expense of LAV-DF.

\begin{table}[!t]
\centering
\caption{Branch diagnostics from the same fused-selected checkpoint.}
\label{tab:branch_diagnostic}
\small
\setlength{\tabcolsep}{4pt}
\renewcommand{\arraystretch}{1.10}
\begin{tabular}{llccc}
\toprule
\textbf{Ratio} & \textbf{Branch}
& \textbf{Macro AUC} & \textbf{Worst AUC} & \textbf{\(\Delta\)AUC} \\
\midrule
\multirow{3}{*}{1:3}
& Shared & 0.9412 & 0.9292 & 0.0239 \\
& Private & 0.7293 & 0.5156 & 0.4274 \\
& Fused & \textbf{0.9561} & \textbf{0.9557} & \textbf{0.0008} \\
\midrule
\multirow{3}{*}{1:9}
& Shared & 0.9453 & 0.9151 & 0.0604 \\
& Private & 0.9371 & 0.8960 & 0.0822 \\
& Fused & \textbf{0.9664} & \textbf{0.9486} & \textbf{0.0358} \\
\bottomrule
\end{tabular}
\end{table}

\subsection{Diagnostic branch analysis}
\label{sec:diagnostic_branch_analysis}

We analyze the shared-only, private-only, and fused branches from the same
fused-selected checkpoint. As shown in Table~\ref{tab:branch_diagnostic}, the
shared branch remains independently discriminative, while the private branch
alone is substantially less competitive, especially under 1:3. The fused
branch achieves the best macro and worst-source AUC in both settings,
supporting the use of the private stream as a residual correction rather than
a standalone detector.

\begin{table}[!t]
\centering
\caption{R19 ablations (FAC/LAV-DF pools: 320/2,880; 3,200 samples/epoch).
Higher Macro/Worst and lower \(\Delta\) AUC are better.}
\label{tab:ablation_summary}
\small
\setlength{\tabcolsep}{4pt}
\renewcommand{\arraystretch}{1.10}
\begin{tabular}{lccc}
\toprule
\textbf{Variant} & \textbf{Macro AUC} & \textbf{Worst AUC} & \textbf{\(\Delta\)AUC} \\
\midrule
Full Ours & \textbf{0.9641} & \textbf{0.9456} & 0.0370 \\
w/o domain & 0.9542 & 0.9390 & 0.0304 \\
w/o gated fusion & 0.9595 & 0.9440 & 0.0310 \\
w/o orth. & 0.9613 & 0.9416 & 0.0394 \\
w/o warm-up & 0.9566 & 0.9363 & 0.0407 \\
\bottomrule
\end{tabular}
\end{table}

\subsection{Ablation studies}
\label{sec:ablation_studies}

As reported in Table \ref{tab:ablation_summary}, ablations use a separate R19 batch whose full model attains a macro AUC of
0.9641, slightly different from the main R19 result of 0.9664. We therefore
compare variants within this batch. We focus on macro and worst-source AUC and
report \(\Delta\)AUC as a source-balance diagnostic.

Removing the domain loss decreases macro AUC from 0.9641 to 0.9542 and
worst-source AUC from 0.9456 to 0.9390. This indicates that explicit source
supervision helps the private stream retain useful source-dependent
information. Removing warm-up also degrades performance, reducing macro AUC
to 0.9566 and worst-source AUC to 0.9363, suggesting that stabilizing the
shared stream before enabling private and fused objectives is beneficial.

The gated residual fusion module also contributes to final performance.
Without gated fusion, macro AUC drops from 0.9641 to 0.9595, supporting the
benefit of injecting source-specific residual evidence into the shared
representation. Removing the orthogonality regularizer leads to a smaller
degradation, suggesting a mild but positive regularization effect.

\subsection{Discussion}
\label{sec:experiment_discussion}
Across both protocols, the shared-private framework improves macro AUC and
reduces the source-domain AUC gap over the unified MSCT pathway, indicating
that a single discrepancy path can be suboptimal for heterogeneous, uneven
source compositions.

The branch analysis clarifies the functional role of each stream. The
shared-only branch remains independently discriminative, indicating useful
cross-source cues. The private-only branch is substantially weaker under R13,
indicating that its main role is to provide source-specific residual correction
rather than standalone prediction. The consistent improvement of the fused
branch over both diagnostic branches supports this interpretation.

The ablations reinforce this view: domain supervision and shared-stream
warm-up support private specialization, while gated fusion converts
source-specific evidence into improvements over the diagnostic branches. The
smaller orthogonality effect indicates that supervised specialization and
feature-level integration are the primary contributors.

\section{Limitations and future work}
\label{sec:limitations_future_work}
This work has several limitations. First, fused and private inference requires
a correct source identity; automatic or soft routing and a shared-only fallback
remain future work. Second, the private stream may encode forgery cues together
with acquisition, compression, preprocessing, or manipulation biases. Because
the regularizers do not guarantee disentanglement, domain probes, gate
statistics, and error analyses are needed. Finally, evaluation is limited to two datasets. Additional sources and broader multi-source baselines are needed, and the number of private experts grows with the source count, although each sample activates only one.

\section{Conclusion}
\label{sec:conclusion}
We presented Pre-MSCT Shared-Private Feature Fusion for known-source imbalanced
multi-source audio-visual deepfake detection. It routes tokens through shared
and source-specific private paths before discrepancy modeling and combines
them through gated residual fusion. Across the evaluated FakeAVCeleb and
LAV-DF protocols, it improves macro-averaged and worst-source performance over
MSCT and MRDF.

\section*{Acknowledgments}
Authors wish to acknowledge Canada Economic Development for Quebec Regions
(CED) and the Ministry of Economy, Innovation and Energy (MEIE) of the
Government of Quebec for their continued support.

{\small
\bibliographystyle{ieee}
\bibliography{egbib}
}
%\newpage
%\onecolumn

%\documentclass[10pt,letterpaper]{article}

%\usepackage{cvpr}
%\usepackage{times}
%\usepackage{graphicx}
%\usepackage{amsmath}
%\usepackage{amssymb}
%\usepackage{booktabs}
%\usepackage{tabularx}
%\usepackage{array}
%\usepackage{float}

%\begin{document}

\section*{Appendix A. Implementation Details}
\label{app:implementation_details}

This appendix summarizes the implementation details of MRDF, MSCT,
and the proposed Pre-MSCT Shared-Private Feature Fusion framework.
The source-sampling protocols are described in the main paper; here,
we provide additional details on preprocessing, model architecture,
and training.

\subsection*{A.1 Data Preprocessing}

Table \ref{tab:app_preprocessing} provides details of the input preprocessing procedures used in our experiments. All methods use the same FakeAVCeleb and LAV-DF data splits and data-loading pipeline, but their model-specific front-ends differ after the loaded audio and visual representations are obtained. Audio is loaded from 16 kHz mono wav files and converted into log-filterbank features. The audio features are stacked with stack order 4, resulting in 104-dimensional frame-level audio features. MSCT and our method project these features with a linear audio token encoder, while MRDF feeds the same loaded audio representation to its AV-HuBERT-style audio front-end.

Video inputs are loaded as frame sequences and normalized. For FakeAVCeleb, training frames are randomly cropped to \(100\times100\) and horizontally flipped, while center cropping to \(100\times100\) is used during evaluation. Audio and video sequences are temporally aligned by padding or truncation in the data pipeline, and MSCT-style models further truncate audio/video token sequences to a common length inside the model. Additional dlib-based face cropping is disabled in the main experiments.

\subsection*{A.2 Encoder and Architecture Configuration}
Table \ref{tab:app_architecture} compares the encoders, backbone networks, and overall architectures of the baseline methods and our proposed framework. MRDF follows an AV-HuBERT-style audio/video feature extractor and fusion encoder. MSCT and our method share the same low-level front-ends: a linear audio token encoder and a WTConv/Res2Net-CBAM visual front-end. Our modification is applied after low-level token encoding, temporal alignment, and classification-token prepending, before the first MSCT block, i.e., before both MSSA and DCA.
\onecolumn
\subsection*{A.3 Training and Hyperparameters}
Table \ref{tab:app_training} summarizes the training configurations used for the baseline methods and our proposed framework. All methods are trained under the controlled source-sampling protocols described in the main paper. Learning rates and training objectives follow the reproduced configurations for each baseline. For checkpoint selection, all methods use validation macro AUC of the main prediction output. For our method, the main output is the fused branch; shared-only and private-only branches are evaluated from the same fused-selected checkpoint for diagnostic analysis. The Ours-specific loss weights in Table~\ref{tab:app_training} correspond to \(\lambda_f\), \(\lambda_p\), \(\lambda_o\), \(\lambda_v\), \(\lambda_\beta\), and \(\lambda_d\), respectively. During the one-epoch warm-up, the private path, fused loss, private loss, and private regularizers are disabled. Sampling is performed with replacement, and source ratios are enforced in expectation at the epoch level, as described in the main paper.
%%% All Tables
%\onecolumn
\begin{table*}[h]
\centering
\caption{Input preprocessing used in our experiments.}
\label{tab:app_preprocessing}
\small
\setlength{\tabcolsep}{5pt}
\renewcommand{\arraystretch}{1.12}
\begin{tabularx}{\textwidth}{p{0.28\textwidth}X}
\toprule
\textbf{Item} & \textbf{Configuration} \\
\midrule
Audio loading
& 16 kHz mono wav \\

Audio feature
& Log-filterbank features with stack order 4 \\

Stacked audio dimension
& 104-dimensional frame-level representation \\

MSCT/Ours audio projection
& Linear audio token encoder \\

MRDF audio front-end
& AV-HuBERT-style audio front-end \\

Video loading
& Frame sequence with normalization \\

FakeAVCeleb train transform
& Random \(100\times100\) crop and horizontal flip \\

FakeAVCeleb eval transform
& Center \(100\times100\) crop \\

Temporal alignment
& Dataset-level padding/truncation; model-level truncation for MSCT-style models \\

Additional dlib crop
& Disabled in the main experiments \\
\bottomrule
\end{tabularx}
\end{table*}
\vspace{-0.25cm}
\begin{table*}[h]
\centering
\caption{Encoder, backbone network and overall architecture comparison.}
\label{tab:app_architecture}
\small
\setlength{\tabcolsep}{4pt}
\renewcommand{\arraystretch}{1.12}
\begin{tabularx}{\textwidth}{p{0.21\textwidth}XXX}
\toprule
\textbf{Item} & \textbf{MRDF} & \textbf{MSCT} & \textbf{Ours} \\
\midrule
Audio front-end
& AV-HuBERT-style
& Linear token encoder
& Same as MSCT \\

Visual front-end
& AV-HuBERT-style
& WTConv/Res2Net-CBAM
& Same as MSCT \\

Feature dimension
& 768
& 256
& 256 \\

Classification token
& First fused token
& Learnable A/V tokens
& Same, before routing \\

Main sequence block
& Fusion encoder
& MSSA + DCA + FFN
& Shared/private MSCT blocks \\

Routing
& None
& None
& Before first MSCT block \\

Private components
& None
& None
& Adapter, expert, classifier, \(\beta_d\) \\

Domain identity
& Not used
& Not used
& Selects private path \\

Main output
& Multimodal classifier
& Multimodal classifier
& Fused classifier \\
\bottomrule
\end{tabularx}
\end{table*}
\vspace{-0.25cm}
\begin{table*}[h]
\centering
\caption{Training configuration.}
\label{tab:app_training}
\small
\setlength{\tabcolsep}{4pt}
\renewcommand{\arraystretch}{1.12}
\begin{tabularx}{\textwidth}{p{0.23\textwidth}XXX}
\toprule
\textbf{Item} & \textbf{MRDF} & \textbf{MSCT} & \textbf{Ours} \\
\midrule
Main losses
& CE / margin
& AV CE + A/V CE + align.
& Shared/fused/private BCE + reg. \\

Unimodal/alignment auxiliary losses
& -
& Enabled
& Disabled \\

Optimizer
& Adam
& Adam
& Adam \\

Learning rate
& \(2{\times}10^{-4}\)
& \(1{\times}10^{-4}\)
& \(1{\times}10^{-4}\) \\

Weight decay
& \(1{\times}10^{-4}\)
& \(1{\times}10^{-4}\)
& \(1{\times}10^{-4}\) \\

Batch setting
& 1, accum. 4
& 1, accum. 4
& 1, accum. 4 \\

Epochs / seed
& 30 / 42
& 30 / 42
& 30 / 42 \\

Warm-up
& -
& -
& 1 shared-only epoch \\

Ours loss weights
& -
& -
& \(0.5,0.1,0.001,0.001,0.001,0.1\) \\

\(\beta_d\) parameters
& -
& -
& \(\beta_{\max}=1.0\), \(\beta_{\rm init}=0.1\) \\

Checkpoint
& Main-output val. macro AUC
& Main-output val. macro AUC
& Main-output val. macro AUC \\
\bottomrule
\end{tabularx}
\end{table*}

%\end{document}

\end{document}